\documentclass[%
 reprint,
 amsmath,amssymb,
 aps,
pra,
]{revtex4-2}
\usepackage{silence}
\usepackage{float}
\usepackage{slashed}
\usepackage{xcolor}
\usepackage{array}
\usepackage{amsmath}
\usepackage{subcaption}
\usepackage{multirow}
\usepackage[english]{babel}
\usepackage{lineno, blindtext}
\usepackage[justification=raggedright,singlelinecheck=false]{caption}
\usepackage{tikz}
\usepackage{adjustbox}
\usepackage{rotating}
\usepackage{rotfloat}
\usepackage{diagbox}
\usepackage[utf8]{inputenc}
\usepackage{graphicx}
\usepackage{dcolumn}
\usepackage{bm}

\usepackage{hyperref}
\usepackage{xcolor}
\hypersetup{
  colorlinks   = true, 
  urlcolor     = red, 
  linkcolor    = blue, 
  citecolor   = blue 
}

\usepackage{threeparttable}

\begin{document}

\preprint{APS/123-QED}

\title{Search for dark matter in the framework of Einstein-Cartan gravity at the International Linear Collider (ILC)}

\author{Hossam Taha$^{1,2}$}
\altaffiliation[Hossam.Taha@bue.edu.eg]{}

\author{El-sayed A. El-dahshan$^{2}$}
\altaffiliation{seldahshan@sci.asu.edu.eg}

\author{S. Elgammal$^3$}
\altaffiliation{Sherif.Elgammal@bue.edu.eg}

\affiliation{\\}
\affiliation{$^{1}$\text{Basic Science Department, Faculty of Engineering, The British University in Egypt}}
\affiliation{$^{2}$\text{Physics Department, Faculty of Science, Ain Shams University}}
\affiliation{$^{3}$\text{Centre for Theoretical Physics, The British University in Egypt}}

\begin{abstract}
This paper investigates the possibility of Dark Matter (DM) fermions production alongside a gauge boson (A$^{\prime}$) using a model based on Einstein-Cartan gravity in an electron-positron linear collider, such as the ILC, that operates at a center-of-mass energy $\sqrt{s} = 500$ GeV with a detector's integrated luminosity of  500  fb$^{-1}$. We used the \texttt{WHIZARD} package as the event generator to simulate the $ e^{+} e^{-}$ interactions that lead to the production of di-muon pairs and missing transverse energy. We specifically are performing this study on a low mass dark gauge boson, M$_{A^\prime}$= 10 GeV, that can subsequently decay into a muon pair (A$^\prime\rightarrow \mu^{+}\mu^{-})$ while aiming to set upper limits on the free parameters' masses of the model, such as the torsion field (ST), if evidence for physics beyond the standard model is not found.

\begin{description}
\item[Keywords]
Dark matter, Dark gauge boson, The International Linear Collider (ILC), Beyond the Standard Model
\end{description}

\end{abstract}

\maketitle


\section{Introduction}
\label{sec:intro}

While the Standard Model of Particle Physics (SM) successfully describes many interactions, including the strong and weak nuclear forces as well as the electromagnetic force, it falls short in accounting for the gravitational force. Additionally, it only accounts for about 5\% of the observable universe. It does not account for dark energy, which is believed to be the cause behind the universe's expansion, nor Dark Matter (DM), our focus in this work, which is suggested as a resolution to the mystery surrounding the unidentified portion of the universe's mass \cite{sm1,sm2,DE}.
DM is a non-luminous form of matter that almost does not interact with known particles in the SM while being strongly supported by numerous cosmological observations \cite{dm1,dm2,dm3}. Exploring DM requires us to go beyond the SM, considering it a low-energy manifestation of broader theories of higher energies that could unify the four fundamental forces \cite{bsm}.

An alternative formulation of general relativity, known as the Einstein-Cartan gravity theory, allows for additional degrees of freedom through the torsion field, that characterizes the torsion aspect of the spacetime \cite{tor1,tor2,tor3,Einstein-Cartan}. In the Einstein-Cartan portal, gravity encoded in the torsion field (ST) can couple through its axial-vector mode to all of the SM fermions. It could provide a way to search for DM by probing the dark gauge boson A$^{\prime}$, which couples to all SM particles through kinetic mixing.
Searching for DM usually involves the production of SM particles plus a large missing transverse energy ($E_T^{miss}$), taken away by the DM particles, with a signature of (SM + $E_T^{miss}$). The search for a DM particle such as the fermion $\chi$ can be conducted by investigating the $E_T^{miss}$ in the final state because the A$^{\prime}$ decays into a SM di-lepton ($A^\prime \rightarrow l^+ l^-)$ that can be detected. However, as the DM particles cannot be detected, they only appear as missing energy. This is illustrated in the signal topology in figure \ref{figure:fig1}.

We propose that linear colliders operating at relatively low energies, such as the International Linear Collider (ILC) operating initially at a center-of-mass energy of $\sqrt s = 500$ GeV and a detector's integrated luminosity of $\mathcal{L} = $ 500  fb$^{-1}$ (with a potential upgrade to $\sqrt s = 1000$ GeV with $\mathcal{L} = $ 1000  fb$^{-1}$) \cite{ilc1,ilc2}, could be well-suited for exploring potential low-mass dark neutral gauge boson candidates. These are beyond what the Large Hadron Collider (LHC) can reach with its current TeV energy scale, and would not have been observed at the Large Electron Positron Collider (LEP), as the considered masses of the studied torsion field are far greater than the $\sim$200 GeV center-of-mass energy of the LEP.

The ATLAS and CMS collaborations have previously searched for a similar dark gauge boson ($Z'$) predicted by the Grand Unified Theory (GUT) and Supersymmetry 
\cite{Extra-Gauge-bosons, gaugeboson1, LR-symmetry11, Super-symmetry12}, 
but found no evidence of its existence in the LHC's RUN II.

The ILC is a future electron-positron linear collider initially designed to work as a Higgs factory to study Higgs bosons. However, it can also be utilized in searching for new particles beyond the SM \cite{bsmilc}. The use of electron-positron collisions instead of the usual proton-proton (p-p) collisions at the LHC provides a cleaner background largely free of complex quark interactions, as the underlying events of p-p collisions will be eliminated. Additionally, it offers higher precision because electrons and positrons are, themselves, elementary particles, which would also make this kind of collision energy efficient as they will use the full center-of-mass energy. Furthermore, the proposed collider uses controllable initial particle energy and beam polarization techniques.

We are utilizing the torsion model to search for DM using monte-carlo event generators to simulate the conditions of the electron-positron interactions at the ILC at $\sqrt s$ = 500 GeV $\mathcal{L} = $ 500  fb$^{-1}$, aiming to set a 95\% confidence limits on the free parameters of this model. 

In this paper, we begin by describing the $U(1)_D$ simplified torsion model based on the Einstein-Cartan gravity theory, which predicts the possible DM fermions production, and the free parameters associated with this model, in addition to a light-dark gauge boson
(M$_{A^\prime} <$ 80 GeV), in the second section (Section \ref{section:model}). Then we move to describe the simulation methods and techniques used in event generation for both the SM background and the model signal in Section \ref{section:simulation}. While in Section \ref{section:Analysis}, we present the plan of the cut-based analysis along with the specific selection criteria. This is followed in Section \ref{section:results} by discussing the results of the analysis and, finally, the summary in Section \ref{section:conclusion}.

\section{EINSTEIN-CARTAN torsion model}
\label{section:model}
\begin{figure}
        \centering
        \includegraphics[width=.8\linewidth]{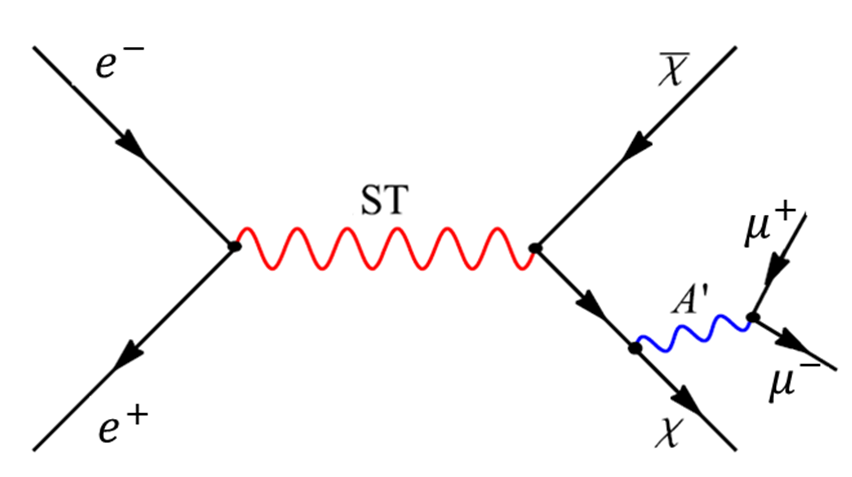}
        \caption{
Signal Topology - An electron interacts with a positron, forming the torsion field (ST) that transforms into two DM particles, while one of them radiates the A$^{\prime}$ via bremsstrahlung.}
  \label{figure:fig1}
    \end{figure}

The simplified torsion model discussed in \cite{Einstein-Cartan} proposes that Einstein-Cartan gravity might serve as a channel for investigating the dark gauge boson (A$^{\prime}$). 
This dark boson could interact with SM particles through the coupling of the torsion field with SM fermions (such as $e^+e^-$) leading to the creation of DM particles ($\chi$  $\bar{\chi}$).
The suggested dark gauge boson $A^\prime_\mu$ can be produced via $e^+e^-$ pair annihilation mediated by the heavy torsion field ($S_\mu$), that represents the axial–vector part of the torsion tensor ($T^{\lambda}_{\mu\nu}$ ), which then
transforms into two DM particles ($\chi$  $\bar{\chi}$). These particles are sufficiently heavy to radiate A$^{\prime}$ via bremsstrahlung, which can subsequently decay into a muon pair ($\mu^+$  $\mu^-$) as shown in figure \ref{figure:fig1}

There are two scenarios in the torsion model. Our focus is on the bremsstrahlung scenario, in which only one of the DM particles radiates the $A^\prime_\mu$ via bremsstrahlung. The other scenario is the cascade scenario, in which both DM particles radiate the A$^{\prime}$. The terms responsible for the interaction between the Dirac fermion $\psi$ and the torsion field in the effective Lagrangian are given by \cite{Einstein-Cartan}: 
\begin{equation}
\bar{\psi}i\gamma^{\mu}(\partial_\mu+ig_{\eta}\gamma^5S_\mu+...)\psi, 
\label{equ1}
\end{equation}
where $\texttt{g}_{\eta}$ represents the coupling of the torsion field to Dirac fermions. The term in the effective Lagrangian describing the coupling between the dark gauge boson $A^\prime_\mu$ and the DM particles, and the coupling of the torsion field to the DM, is given by \cite{Einstein-Cartan}: 
\begin{equation}
\bar{\chi}(i\gamma^{\mu}D_\mu-M_\chi)\chi,
\label{equ2}
\end{equation}
where $D_\mu=\partial_\mu+ig_{\eta}\gamma^5S_\mu+ig_D A^\prime_\mu, M_\chi$ represents the mass of the DM particle and $\texttt{g}_{D}$ is the coupling of the dark gauge boson to the DM particle.

The neutral dark gauge boson (A$^{\prime}$) decays to two SM fermions. In our case, we choose the muonic decay of A$^{\prime}$. The highest significant branching ratio of $A^\prime\rightarrow\mu^+\mu^-$ could be reached if the following mass assumption is satisfied \cite{Einstein-Cartan}:
\begin{equation}
M_{A^\prime}<2M_\chi.
\label{equ3}
\end{equation}

The model's free parameters include the masses of the torsion field ($M_{ST}$), the dark gauge boson ($M_{A^\prime}$ ), the DM fermion ($M_\chi$) and the coupling
constants ($\texttt{g}_{\eta}$ and $\texttt{g}_{D}$).
Now, because this model was theoretically investigating the DM production through only p-p collisions, we had to tune some of the parameters within acceptable limits to suit our desired $e^+e^-$ colliders.
Thus, $\texttt{g}_{\eta}$ was set to 0.25 for all leptons, as it is considered a free parameter in the case of a dynamical ST that varies around the classical value under quantum fluctuations. In contrast, for quarks, $\texttt{g}_{\eta}$ was set to 0 to avoid possible constraints from past LHC resonance searches. The coupling constant $\texttt{g}_{D}$ was set to 1.2, as taken from \cite{Einstein-Cartan}. Such a choice for these parameters is to achieve the highest possible cross-section multiplied by the branching ratio.

In the search for new heavy resonances in the dilepton channel, the mass range of 0.6 - 5.15 TeV has been excluded by the CMS experiment \cite{cms}, while the ATLAS experiment has ruled out the mass range between 0.6 - 5.1 TeV \cite{ATLAS}. Additionally, the CMS collaboration has conducted a search for a narrow resonance lighter than 200 GeV in \cite{zprimeCMSlowmass}. This search has excluded the resonance mass range of 11.5–45.0 GeV, prompting us to choose an A$^{\prime}$ mass lower than 11.5 GeV.
Since we are interested in studying the possible production of DM particles at linear colliders like the ILC, along with light neutral dark gauge boson, we have fixed the mass of A$^{\prime}$ to be M$_{A^\prime}$ = 10 GeV to satisfy the mass condition given in equation (\ref{equ3}). The usual signature of this process is composed of a pair of oppositely charged muons from the A$^{\prime}$ decay and a large $E_T^{miss}$ resulting from the non-decaying DM fermion $\chi$, serving as a clean channel to SM backgrounds. This gives our studied events a topology of $\mu^+\mu^-+E_T^{miss}$.

\section{Simulated Signal samples and SM backgrounds}
\label{section:simulation}

The SM background processes that produce muon pairs within the signal region are Drell-Yan ($DY \rightarrow \mu^+\mu^-$) production, top quark pairs production ($t\bar{t} \rightarrow \mu^+\mu^- + 2b + 2\nu$) and production of diboson ($W^+W^- \rightarrow \mu^+\mu^- + 2\nu$, $ZZ \rightarrow \mu^+\mu^- + 2\nu$, and $ZZ \rightarrow 4\mu$). We used \texttt{WHIZARD}~v2.8.5 \cite{whiz}, a general-purpose matrix element event generator, to generate the model signal events. The files used in implementing the model into \texttt{WHIZARD} were requested from the author of \cite{Einstein-Cartan}. The hadronization process was done with \texttt{PYTHIA v6} \cite{pyth}.
As for the simulation of the ILC detector (the ILD), the reconstruction processes, and the simulation of read-out system response (digitization), we used \texttt{DELPHES}~\cite{delph}, for a fast detector simulation of the ILD. The events were generated from electron-positron collisions at the ILC with 500 GeV center-of-mass energy, and an integrated luminosity of $ 500  fb^{-1}$, with polarized degrees of $P_{e^-} = 0.8$ for the electron beams, and $P_{e^+} = -0.3$ for the positron beams which corresponds to the circumstances discussed in \cite{ilc1,ilc2}.

\begin{figure}[!ht]
        \centering
        \includegraphics[height=0.24\textwidth]{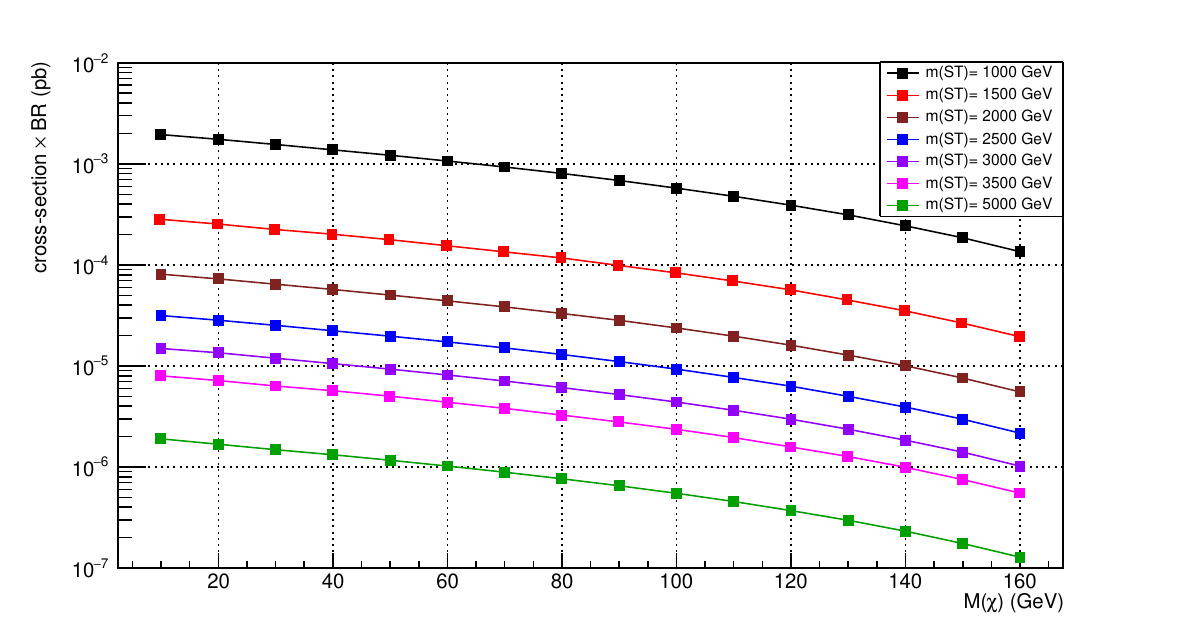}
        \caption{Plot of the cross-section (in log scale) on the y-axis vs M$_\chi$ on the x-axis at different masses of the torsion field. The statistical uncertainty of the calculation is about 0.2\%}
  \label{figure:fig2}
    \end{figure}
\begin{figure}[!ht]
        \centering
        \includegraphics[height=0.24\textwidth]{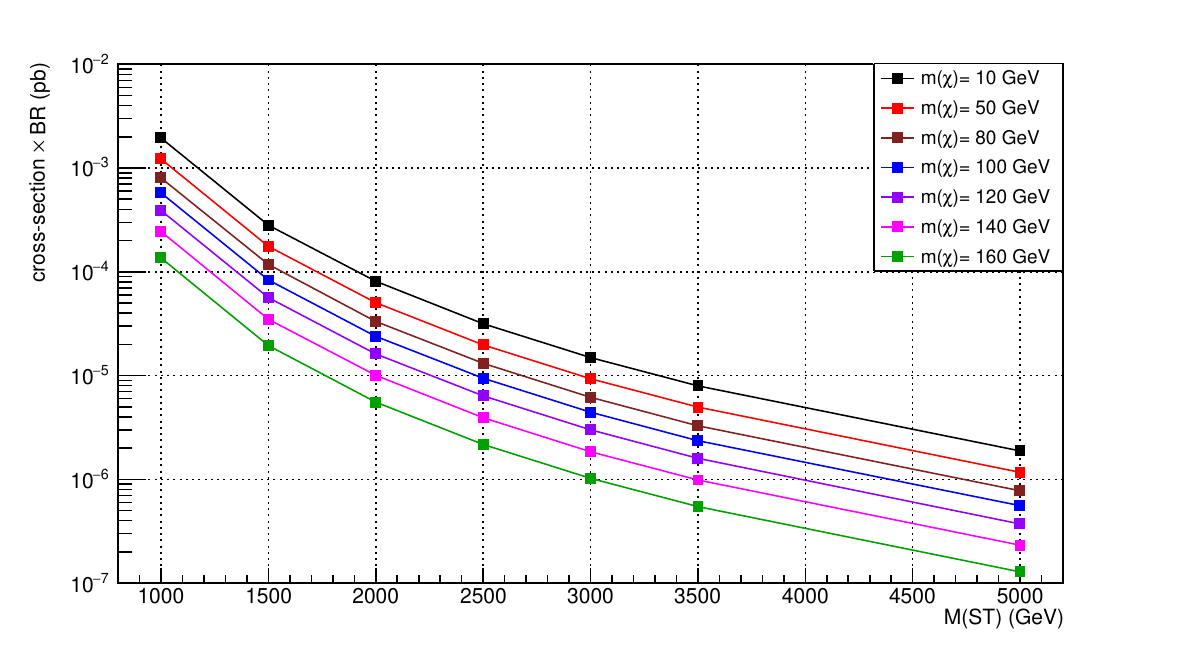}
        \caption{Plot of the cross-section (in log scale) on the y-axis at different masses of the DM ($\chi$) vs the torsion field on the x-axis. The statistical uncertainty of the calculation is about 0.2\%}
  \label{figure:fig3}
    \end{figure}
For different values of the DM particle ($\chi$) and torsion field ($ST$) masses, the measurements of cross-section times the branching ratio ($\times$ BR) were calculated along with the errors which is the 1-$\sigma$ theoretical uncertainty on the cross-section measurement computed on a statistical basis and plotted in log scale against the masses of $\chi$ and ST as shown in figures \ref{figure:fig2}, \ref{figure:fig3}.
The used Monte Carlo samples (signal and background events) in our work were calculated at leading order (LO), as well as for their corresponding cross-sections, and are indicated in table \ref{table:tab3}. Thus, the contributions of our processes (SM background and signal samples) have been determined using the Monte Carlo simulations, normalized to their cross-sections, and an integrated luminosity of 500 fb$^{-1}$.
\begin{table*}
\centering
\begin{tabular} {|l|l|c|c|c|}
\hline
\hspace{0.2cm} Process \hspace{0.2cm} & \hspace{0.2cm} Decay channel \hspace{0.2cm}  & \hspace{0.3cm} Generator \hspace{0.3cm} & \hspace{0.3cm} $\sigma \times \mathrm{BR}$ (fb) \hspace{0.3cm} & \hspace{0.3cm} Order \hspace{0.3cm} \\
\hline
\hline
$DY$  & $\mu^+\mu^-$ & \texttt{WHIZARD} & 1850 $\pm$ 70  & LO \\
\hline
$t\bar{t}$ & $\mu^+\mu^- + 2b + 2\nu$  & \texttt{WHIZARD} & 10.36 $\pm$ 0.02 & LO \\
\hline
WW  & $\mu^+\mu^- + 2\nu$ & \texttt{WHIZARD} & 232.5 $\pm$ 0.6 & LO \\
\hline
ZZ & $\mu^+\mu^- + 2\nu$ & \texttt{WHIZARD} & 3.72 $\pm$ 0.01 & LO \\
\hline
ZZ  & $4\mu$ & \texttt{WHIZARD} & 0.470 $\pm$ 0.002 & LO\\
\hline
\end{tabular}
\caption{SM background processes that mimic our signal, with their corresponding event generator, cross-sections multiplied by the branching ratio for each process, and the order of calculations.}
\label{table:tab3}
\end{table*}

In simulating the SM processes which include muons and/or missing energy in the final state (accounting for the neutrinos that are not detected) that can interfere with the signal events, we employed the \texttt{WHIZARD} package, given that the ILC is not operational yet, and used its built-in interface with \texttt{PYTHIA} to model showering and hadronization.

\begin{table*}
\centering
\renewcommand{\arraystretch}{1.5} 
\begin{tabular} {|c|c|c|}
\hline
Step & Variable & Requirements \\
\hline
\hline

\multirow{3}{*}{Pre-selection}

& $p^\mu_T \ (GeV) $  & $> 10$ \\ \cline{2-3}

& $|\eta^\mu| \ (rad) $ & $< 2.4$ \\ \cline{2-3}

& IsolationVar & $< 0.1$ \\ \cline{2-3}

\hline
\hline

\multirow{6}{*}{Tight selection} 

& $|P^{\mu^+\mu^-} -{E}_T^{\text{miss}}|/P^{\mu^+\mu^-} \  $ & $< 0.1$ \\ \cline{2-3}

& Mass window (GeV) & $ M_{A^\prime} \times 0.5 < M_{\mu^+\mu^-} \ < M_{A^\prime} \times 1.5 $ \\ \cline{2-3}

& $\Delta\phi_{\mu^+\mu^-,\vec{E}{_T^{miss}}} \ (rad) $ & $ > 2.8$ \\ \cline{2-3}
&$\Delta R_{\mu^+\mu^-}  $ & $< 0.8$ \\ \cline{2-3}

& Cos($\theta_{3D}$) & $<-0.8$ \\ \cline{2-3}

& number of b-tagged jets & $< 1$ \\ \hline

\end{tabular}
\caption{Summary of the final event selection used in the analysis.}
\label{cuts}
\end{table*}

\section{Events selection and analysis strategy}
\label{section:Analysis}

The event selection for this analysis is designed to reconstruct a final state of muons with low transverse momentum ($p_T$) in association with $E_T^{miss}$, accounting for the DM candidates ($\chi \Bar{\chi}$). The selection criteria are applied in the form of cuts on different kinematic parameters, as detailed in table \ref{cuts}. 
Where $\eta^\mu$ represents the pseudo-rapidity for the di-muon. “IsolationVar” refers to an isolation cut in DELPHES, used to reject muons produced inside jets. In this cut, the scalar $p_T$ sum of all tracks within a cone of $\Delta R = 0.5$ around the muon candidate, excluding the muon candidate itself, must not exceed 10\% of the muon's $p_T$. Thus, each event is selected to include two oppositely charged muons.  After applying the pre-selection cuts, at which each event must contain exactly two muons with opposite charges, the tight selection is optimized for DM signals to distinguish them from the SM background, thereby obtaining the best-expected limit.

Figure \ref{figure:fig5} shows the $E_T^{miss}$ distributions for events passing the pre-selection cuts listed in table \ref{cuts}, which are stacked on each other. The light green histogram represents the $t\bar{t}$ background, the yellow histogram stands for the vector boson pair backgrounds (ZZ and WW), and the cyan histogram represents the Drell-Yan background.

The non-filled histogram represents the bremsstrahlung scenario within the framework of the Einstein-Cartan gravity, which has been generated with fixed values for the neutral light gauge boson A${^\prime}$ and DM ${\chi}$ masses ($M\chi = 10$ GeV, $M_{A^\prime} = 10$ GeV) while varying the torsion field mass, which is represented by different colored histograms, and overlaid on the stacked histograms. 
As shown in figure \ref{figure:fig5}, the SM backgrounds dominate the signal samples. Thus, applying a tighter selection is required to discriminate our signal from the SM backgrounds. Such selection is based on the following six variables:

\begin{figure}[H]
        \centering
        \includegraphics[width=1\linewidth]{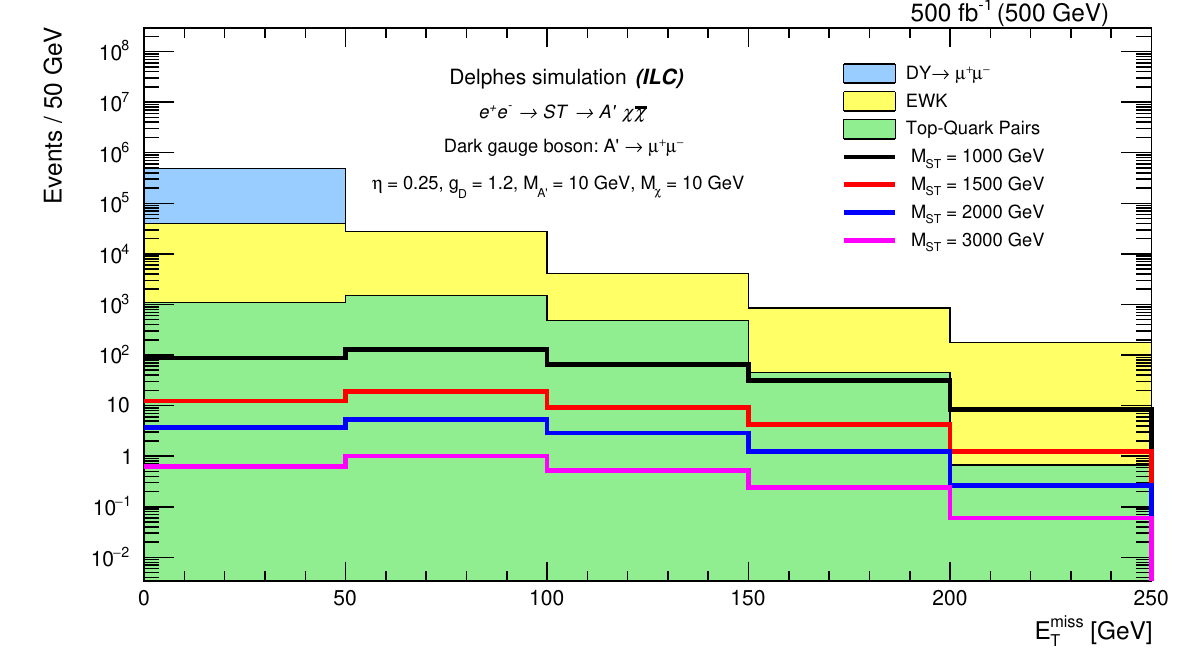}
        \caption{The missing transverse energy after applying the pre-selection cuts listed in table \ref{cuts} for both the SM background and the torsion model at different ST mass points (M$_{ST}$= 1000, 1500, 2000, 3000 GeV) and at a DM mass of M$_\chi$ = 10 GeV, and M$_{A^\prime}$ = 10 GeV.}
  \label{figure:fig5}
    \end{figure}

\begin{figure*}
    \centering
    \begin{subfigure}[b]{0.475\textwidth}
        \centering
        \includegraphics[width=\textwidth]{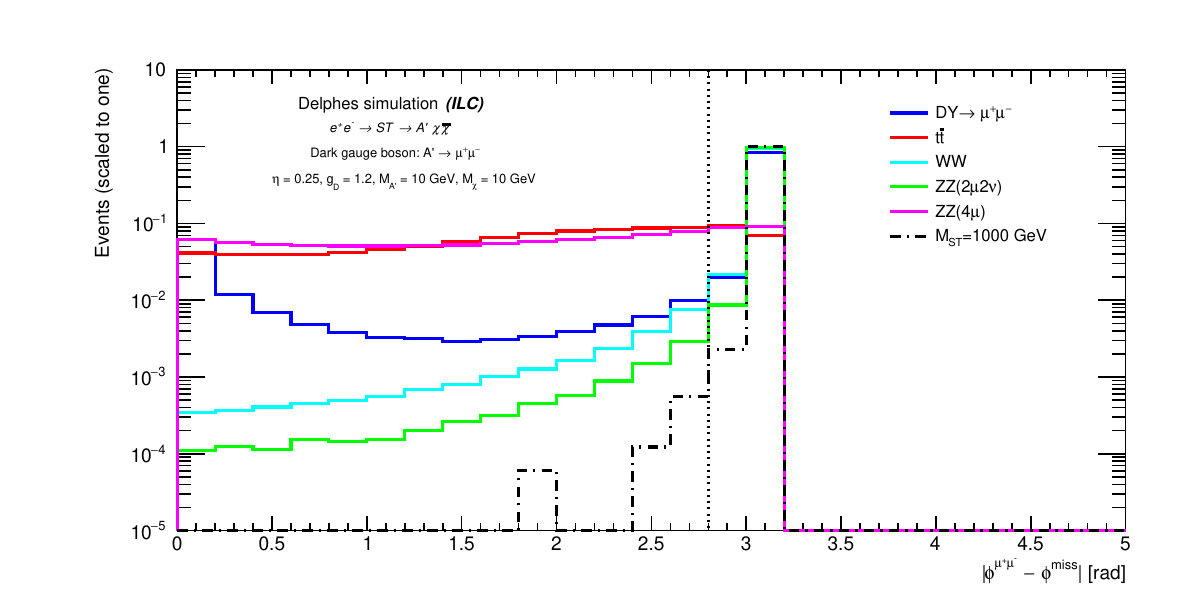}
        \caption{$\Delta\phi_{\mu^+\mu^-,\vec{E}{_T^{miss}}}$}    
        \label{cut1}
    \end{subfigure}
    \hfill
    \begin{subfigure}[b]{0.475\textwidth}  
        \centering 
        \includegraphics[width=\textwidth]{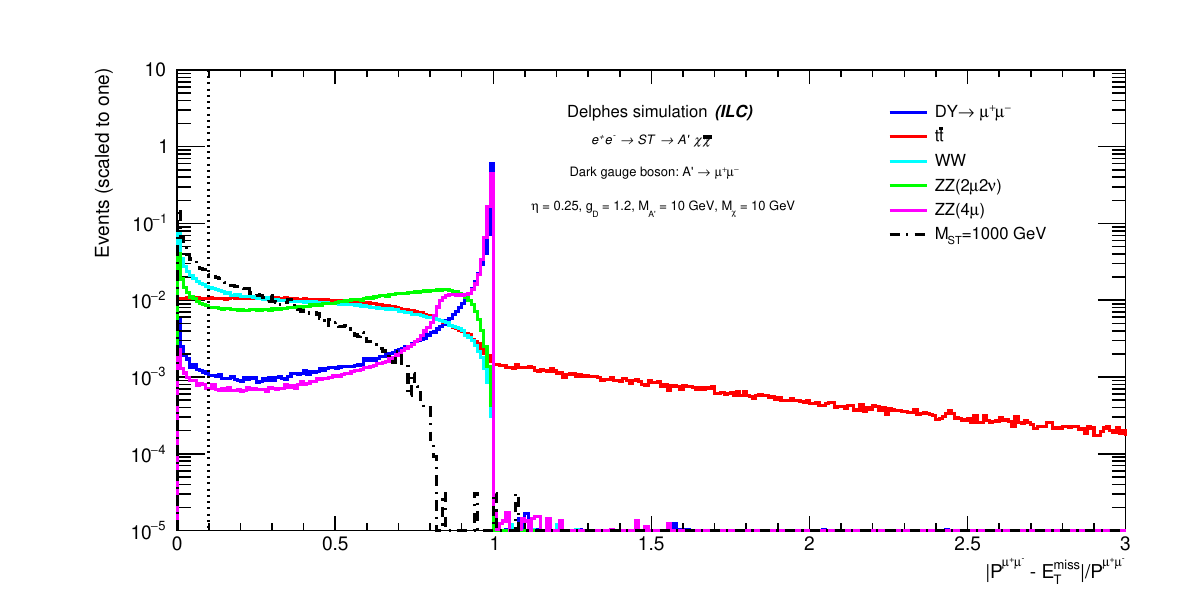}
        \caption{$|P^{\mu^+\mu^-} -{E}_T^{\text{miss}}|/P^{\mu^+\mu^-}$}    
        \label{cut2}
    \end{subfigure}

    \vskip\baselineskip 

    \begin{subfigure}[b]{0.475\textwidth}   
        \centering 
        \includegraphics[width=\textwidth]{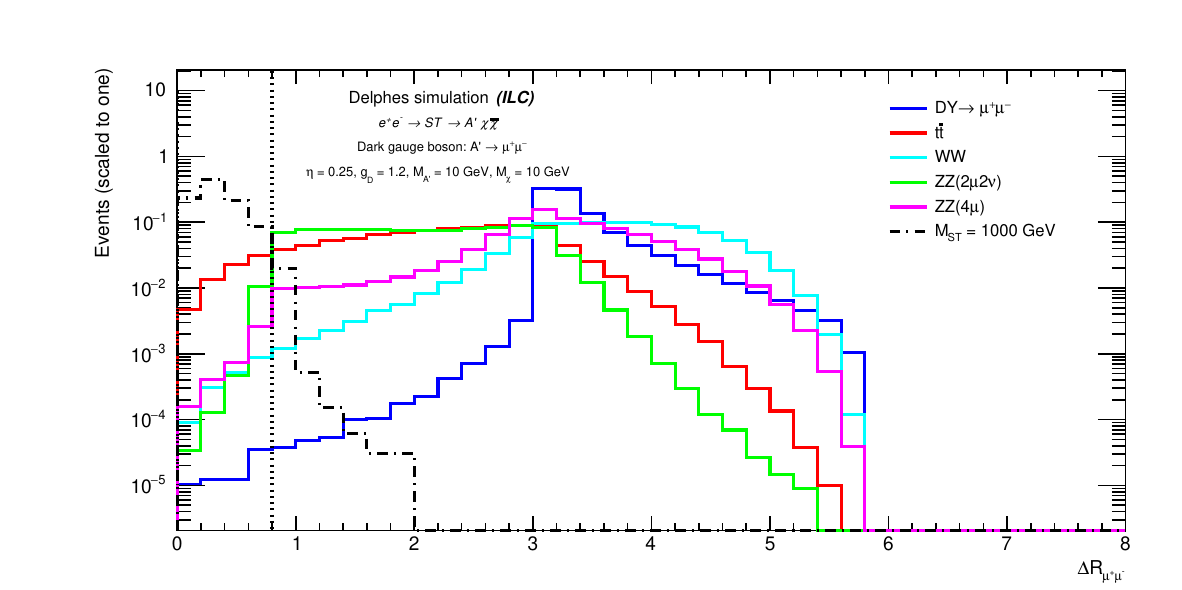}
        \caption{$\Delta R_{\mu^+\mu^-}$}    
        \label{cut3}
    \end{subfigure}
    \hfill
    \begin{subfigure}[b]{0.475\textwidth}   
        \centering 
        \includegraphics[width=\textwidth]{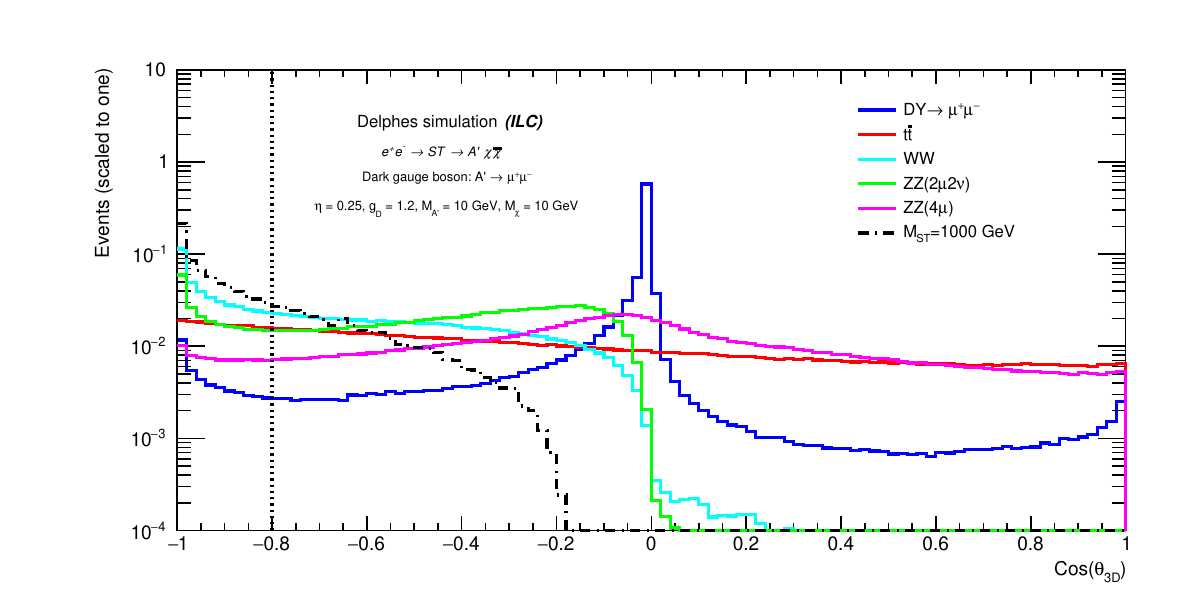}
        \caption{Cos($\theta_{3D}$)}    
        \label{cut4}
    \end{subfigure}
    \hfill
    \begin{subfigure}[b]{0.475\textwidth}   
        \centering 
        \includegraphics[width=\textwidth]{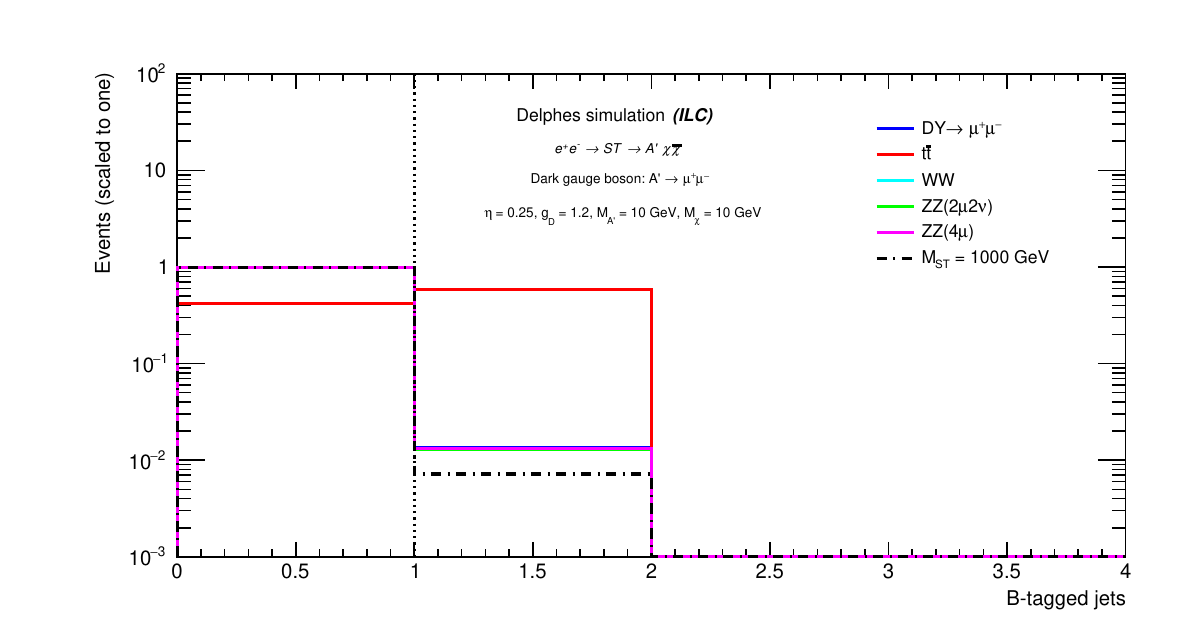}
        \caption{formed b-tagged jets}
        \label{cut5}
    \end{subfigure}

    \caption{The variables employed in the tight cuts. The histograms are normalized to unity to emphasize the overall characteristics. A vertical black dotted line indicates the cut-off value.}
    \label{figure:cuts}
\end{figure*}

\noindent (1) $|P^{\mu^+\mu^-} -{E}_T^{\text{miss}}|/P^{\mu^+\mu^-}$, which is the relative difference between the magnitudes of di-muon's total momentum $P^{\mu^+\mu^-}$, and the missing transverse energy ${E}_T^{\text{miss}}$. \\ 
(2) The mass of the di-muon system (M$_{\mu^+\mu^-}$) is required to be within $0.5 \times M_{A^{\prime}} < M_{\mu^+\mu^-} < 1.5 \times M_{A^{\prime}}$ to preserve consistency with leptons produced by the A$^{\prime}$ boson decay. Having M$_{A^\prime}$= 10 GeV, the mass window is then set to be in the range of 5 to 15 GeV. \\
(3) The azimuthal angle difference between the di-muon system and the missing transverse energy $\Delta\phi_{\mu^+\mu^-,\vec{E}{_T^{miss}}}$.\\  
(4) $\Delta R_{\mu^+\mu^-}$, the angular distance between the two muons. \\
(5) Cosine of the 3D angle ($\theta_{3D}$) between the missing transverse energy vector and the di-muon system vector.\\
(6) The B-tagged jets, which allows us to identify the produced bottom quark jets that are found in the $t\bar{t}$ background. It utilizes a Boolean function that returns 0 (the first bin) if no jets are formed and 1 (the second bin) if jets are formed. \\
These cuts (figures; \ref{cut1}, \ref{cut2}, \ref{cut3}, \ref{cut4}, and \ref{cut5}) succeed in suppressing the SM background as seen in figure \ref{figure:fig9}.

\section{Results}
\label{section:results}

\begin{figure}[ht]
        \centering
        \includegraphics[width=1\linewidth]{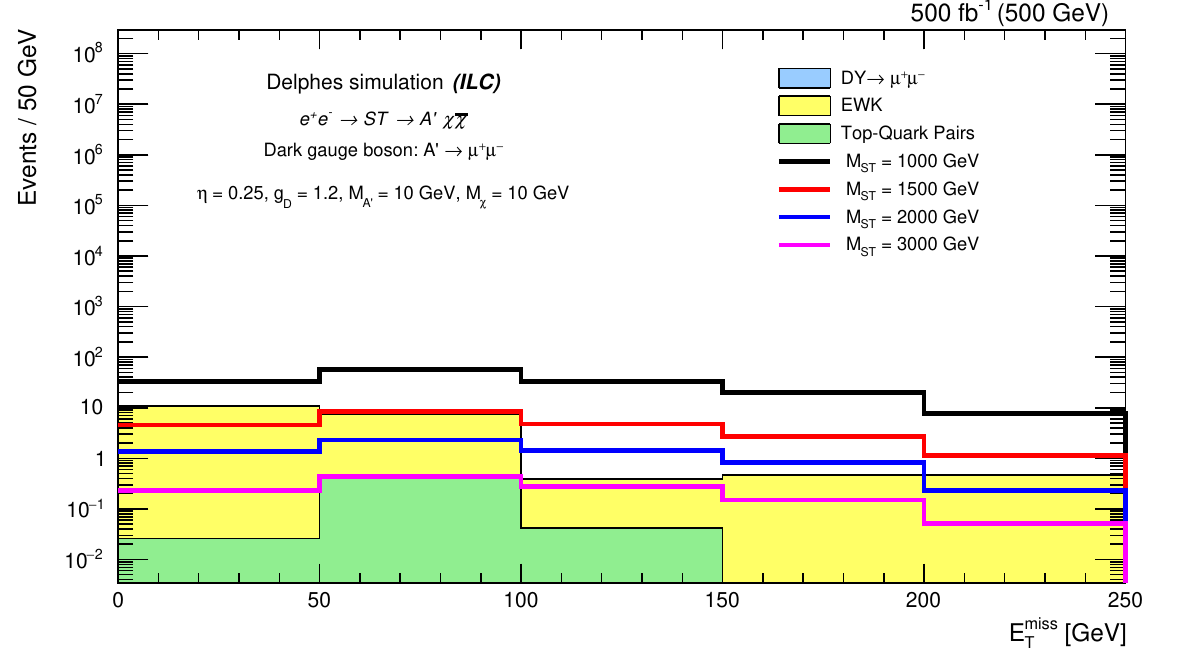}
        \caption{The missing transverse energy after applying the final cuts listed in table \ref{cuts}. The torsion field signal at M$_{ST} =$ 1000, 1500, 2000 and 
        3000 GeV are superimposed for M$_\chi$ = 10 GeV, and M$_{A^\prime}$ = 10 GeV.}
  \label{figure:fig9}
    \end{figure}

In the di-muon channel, we used a shape-based analysis. The missing transverse energy distribution is an effective discriminant variable. The signal processes lead to relatively larger ${E^{miss}_T}$ values than those of the SM backgrounds. Figure \ref{figure:fig9} shows the missing transverse energy distribution, after applying the final event selection, highlighting a significant decrease in the SM background processes.  

The number of di-muon events passing the pre-selection and the final selection are outlined in table \ref{table:tab5} for each SM background process and the signal samples with torsion mass $M_{ST}$ of 1000, 1500 GeV, $M_{\chi}$ = 10 GeV, and $M_{A^\prime}$ = 10 GeV, with an integrated luminosity of 500 fb$^{-1}$.
Adding the tight selection cuts (listed in table \ref{cuts}) to the pre-selection resulted in almost complete suppression of the DY process, and the other SM backgrounds were reduced by roughly three orders of magnitude. 
An ad-hoc flat 10\% uncertainty is applied to cover all possible systematic effects.
The total uncertainties, including both the statistical and systematic components, have been taken into account for the simulated signal and background samples.
\begin{table*}
\centering
\renewcommand{\arraystretch}{1.2} 
\begin{tabular} {|c|c|c|}
\hline
Process  & \hspace{0.3cm} No. of events passing pre-selection \hspace{0.3cm}  & \hspace{0.5cm} No. of events passing final selection \hspace{0.5cm} \\
\hline
\hline
$DY$  & 449000 $\pm$ 45000 & 0.0 \\
\hline
$t\bar{t}$ & 3145 $\pm$ 320  & 0.5 $\pm$ 0.7  \\
\hline
WW  &  68000 $\pm$ 6800  & 18.8 $\pm$ 4.7  \\
\hline
ZZ($4\mu$) & 2520 $\pm$ 260 & 0.0 \\ 
\hline
ZZ($2\mu 2\nu$) & 230 $\pm$ 28 & 0.0  \\
\hline
Total background & 522895 $\pm$ 52408 & 19.5 $\pm$ 4.8 \\
\hline
Signal: $M_{ST}$= 1000,  $M_\chi$=10 (GeV) & 319 $\pm$ 37 & 150 $\pm$ 19.4 \\
\hline
Signal: $M_{ST}$= 1500,  $M_\chi$=10 (GeV) & 46 $\pm$ 8 & 21.6 $\pm$ 5.1 \\
\hline
\end{tabular}
\caption{Number of di-muon events passing pre-selection and final selection for both the SM background and the model signals (at $\sqrt s$ = 500 GeV, $\mathcal{L} = $ 500  fb$^{-1}$).
The total uncertainties, including both the statistical and systematic components, have been taken into account for the simulated signal and background samples.}
\label{table:tab5}
\end{table*}

\begin{figure}[ht]
        \centering
        \includegraphics[width=1.1\linewidth]{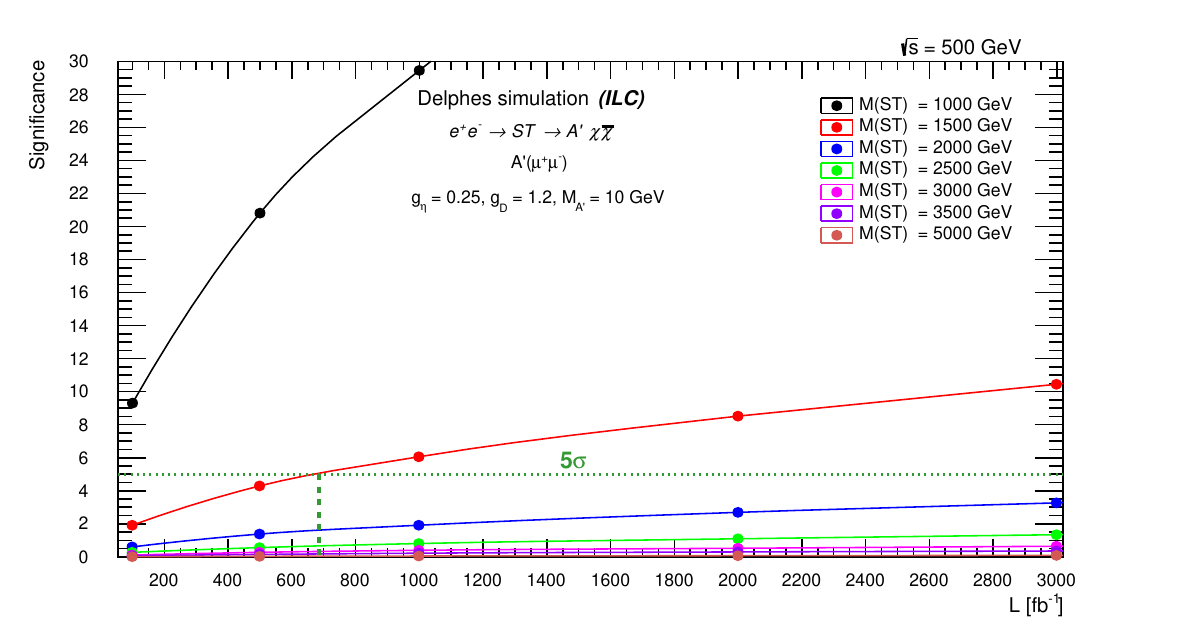}
        \caption{The statistical significance plot versus the ILD integrated luminosity after the final event selection. The green horizontal dotted line points to the statistical $5\sigma$ value.}
  \label{figure:sig}
    \end{figure}

We used the Asimov formula to calculate the statistical significance, $S$, as described in \cite{sig}

\begin{equation}
    S = \sqrt{2 \times \left(
    \left(N_s+N_b\right)\log\left(1+\frac{N_s}{N_b}\right) - N_s\right)},
\end{equation}
where $N_s, N_b$ are the numbers of signal and total SM background events passing certain selection criteria.
We also plotted the statistical significance against the detector's luminosity (figure \ref{figure:sig}) to identify the values of M$_{ST}$ that lead to the desired 5$\sigma$ discovery 
at the final selection stage. Figure 
\ref{figure:sig} underscores the critical role of the tight selection criteria outlined in table \ref{cuts} in distinguishing between the signal and the SM backgrounds to achieve a 5$\sigma$ discovery. It also indicates that for the $M_{ST} =$ 1500 GeV mass point, a 5$\sigma$ discovery could be attained with a luminosity of 685 fb$^{-1}$.

In analyzing our results statistically, we used the profile likelihood method and the modified frequentist construction CLs \cite{CL1,CL2} used in the asymptotic approximation \cite{sig} to set exclusion limits at 95\% confidence level on the product of signal cross-sections and branching ratio Br($A^\prime \rightarrow \mu^+\mu^-$). 

\begin{figure}[!ht]
	\centering
    \begin{subfigure}{0.50\textwidth}
		\includegraphics[width=\textwidth]{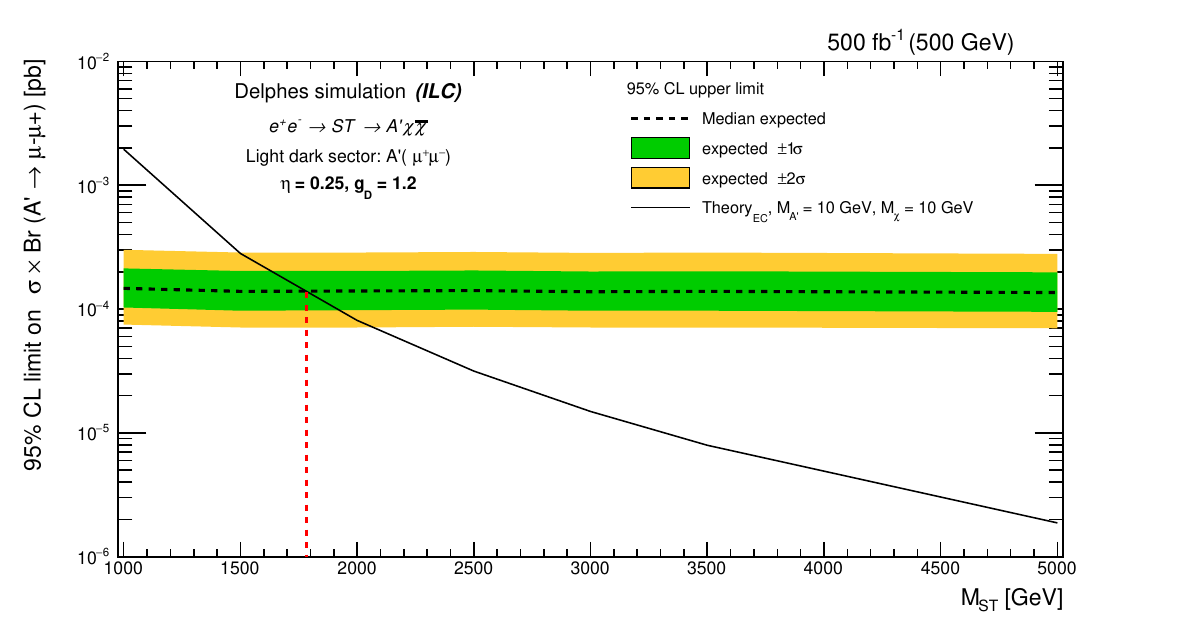}
		\caption{For M$\chi$= 10 GeV, the intersection with the theoretical line occurred exactly at $M_{ST}$ = 1787 GeV.}
        \label{limit1}  
	\end{subfigure}
	\begin{subfigure}{0.50\textwidth}
		\includegraphics[width=\textwidth]{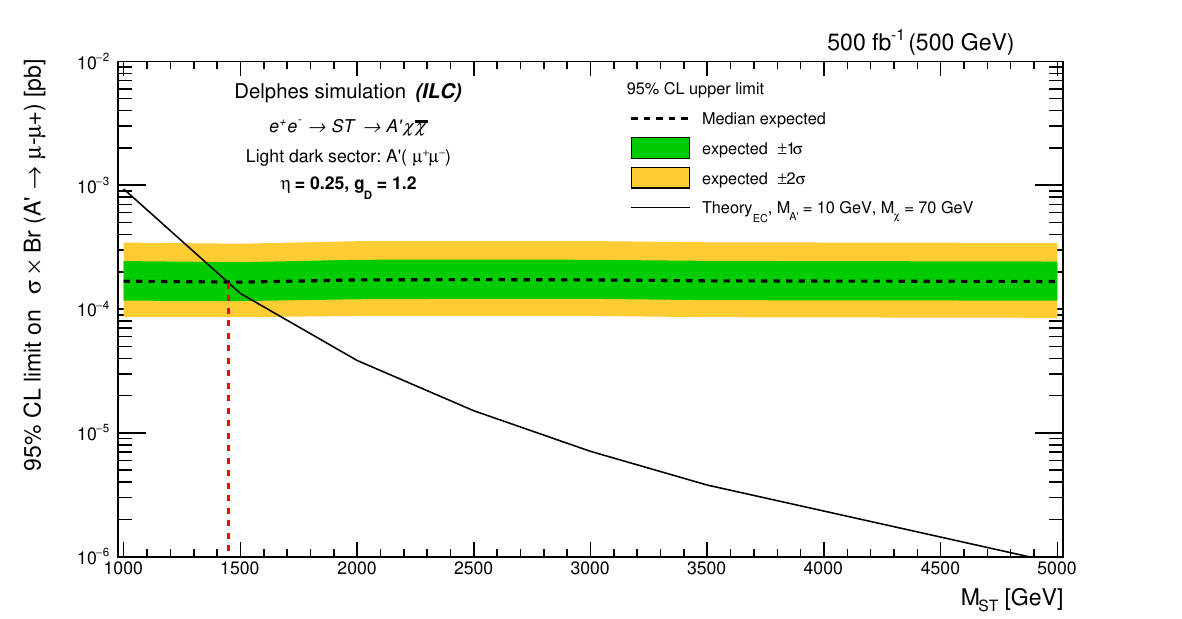}
		\caption{For M$\chi$= 70 GeV, the intersection with the theoretical line occurred exactly at $M_{ST}$ = 1444 GeV.}
        \label{limit2}
	\end{subfigure}
	\begin{subfigure}{0.50\textwidth}
		\includegraphics[width=\textwidth]{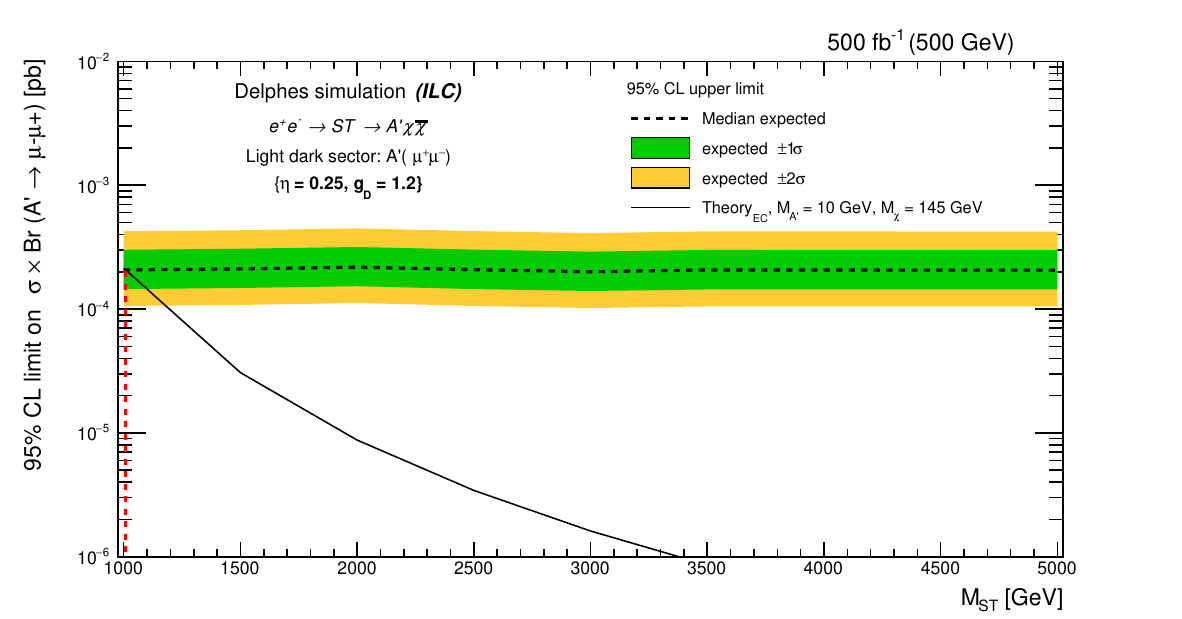}
		\caption{For M$\chi$= 145 GeV, the last possible intersection with the theoretical line occurred exactly at $M_{ST}$ = 1008 GeV.}
        \label{limit3}  
	\end{subfigure}
	\caption{95\% CL upper limits on the cross-section multiplied by the branching ratio (expected), as a function of the mediator's mass (M$_{ST}$), considering the muonic decay of A$^\prime$. The black line represents the torsion model at different $\chi$ mass points $M_{\chi} =$ 10 GeV (\ref{limit1}), 70 GeV (\ref{limit2}), and 145 GeV (\ref{limit3}), with M$_{A^\prime}$ = 10 GeV.
    A vertical red dotted line indicates the upper limit value.}
\label{figure:limits}
\end{figure}

Figure \ref{figure:limits} displays the 95\% upper limit on the cross-section times the branching ratio as a function of the mass of the torsion field M$_{ST}$ within the framework of the Einstein-Cartan gravity model.
In the context of muonic decay of $A^\prime$ with coupling constant $\texttt{g}_{\eta}$ = 0.25 and $\texttt{g}_{D}$ = 1.2, and DM mass $M_\chi$ = 10 (\ref{limit1}), 70 (\ref{limit2}), 145 (\ref{limit3}) GeV for M$_{A^\prime}$ = 10 GeV. We have observed from figure \ref{figure:limits} that the exclusion of torsion field (ST) production applies to the mass range below 1787 in \ref{limit1}, 1444 in \ref{limit2}, and 1008 GeV in \ref{limit3}, as the expected median indicates. 

\begin{figure}
        \centering
        \includegraphics[width=1.1\linewidth]{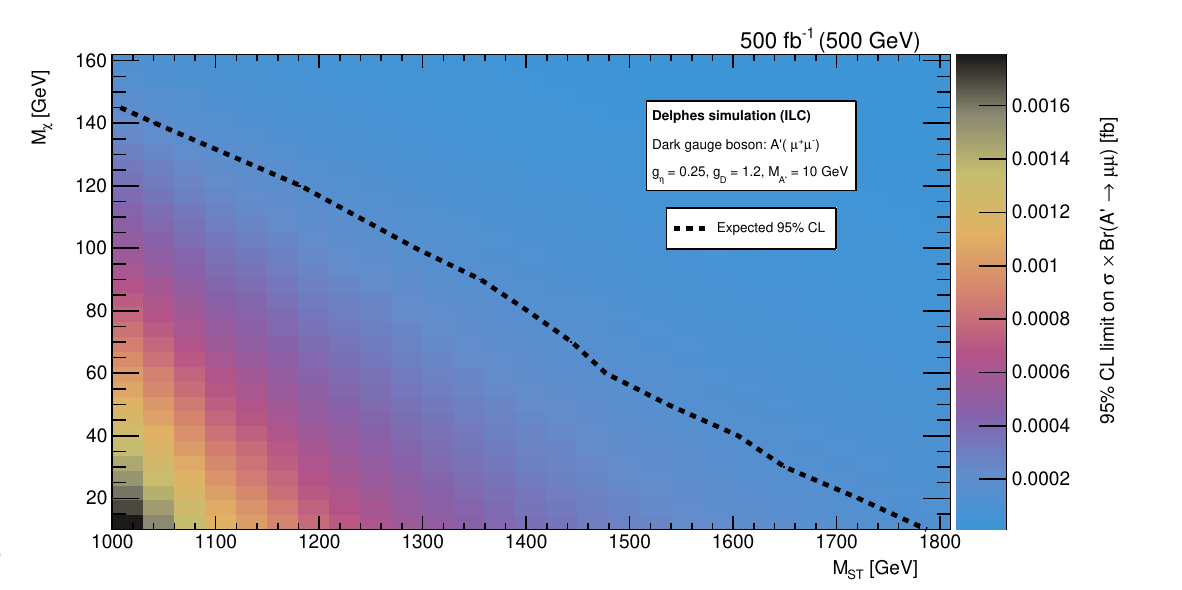}
        \caption{The 95\% CL limits set on the product of cross-section multiplied by the branching ratios for a variety of values of the model's parameters (M$_{ST}$, M$_\chi$). The region below the black dots is expected to be excluded.}
  \label{figure:2Dlimit}
    \end{figure}

In figure \ref{figure:2Dlimit}, the simplified model illustrates the cross-section multiplied by the branching ratio limit. 
This graph pertains to the mediator’s mass, $M_{ST}$, and the DM mass, $M_\chi$. The region below the 95\% expected limit (depicted by the black dotted line) is excluded. 
The findings from the collective signal regions exclude anticipated values falling within $1008 < M_{ST} < 1787$ GeV.

\section{Summary}
\label{section:conclusion}

A search for the production of DM fermions along with a light-dark gauge boson A$^{\prime}$ has been conducted in the framework of Einstein-Cartan gravity.
We used MC samples in this search generated using the \texttt{WHIZARD} package to simulate electron-positron collisions at $\sqrt{s} = $ 500 GeV, with an integrated luminosity of 500 fb$^{-1}$.
Results from the muonic decay mode of A$^{\prime}$ and the statistical and systematic combination of uncertainties are presented.
We used a benchmark signal to test the bremsstrahlung scenario, trying various torsion field (ST) masses on the propagator. We kept the coupling constants $\texttt{g}_{\eta} = 0.25$ and $\texttt{g}_{D} = 1.2$ fixed to attain the highest possible cross-section times branching ratio, with polarized degrees of $P_{e^-} = 0.8$ for the electron beams, and $P_{e^+} = -0.3$ for the positron beams.

By implementing the specified cuts (detailed in table \ref{cuts}), a significant drop in the SM backgrounds was noticed without affecting the strength of the signal. This, in turn, enabled us to achieve a $5\sigma$ discovery for $M_{ST} =$ 1000 and 1500 GeV.
If the mass value, $M_{ST}$, is above 1500 GeV, achieving a 5$\sigma$ discovery would be challenging due to the shallow cross-section (multiplied by the branching ratio). 
This situation may necessitate the implementation of additional new cuts to reduce the SM backgrounds such as $t\bar{t}$ and WW.

If the ILC does not detect this signal, we can establish upper limits on the masses of the torsion field ($M_{ST}$) and DM $(M_{\chi})$ at the 95\% confidence level for the decay of A$^{\prime}$ in the charged muonic channel.
We have set limits that exclude the $M_{ST}$ range from 1008 to 1787 GeV and also exclude the ranges between 10 - 145 GeV for the DM mass ($M_{\chi}$). Further limits can be implemented on other parameters, such as the mass of A$^{\prime}$, in future work.
\begin{acknowledgments}
We want to thank Cao H. Nam, the author of \cite{Einstein-Cartan}, for sharing the UFO file of the torsion model with us that was used in \texttt{WHIZARD} for generating signal events.
\end{acknowledgments}

\textbf{Data Availability Statement:} This manuscript has no associated data or the data will not be deposited. 
\nocite{*}



\begin{thebibliography}{6}


\bibitem{sm1} M. K. Gaillard, P. D. Grannis, F. J. Sciulli, ``The standard model of particle physics", Reviews of Modern Physics, 71(2), S96–S111, 1999. \textcolor{blue}{https://doi.org/10.1103/revmodphys.71.s96}

\bibitem{sm2} A. Crivellin, B. Mellado, ``Anomalies in particle physics and their implications for physics beyond the standard model", Nat Rev Phys 6, 294–309, 2024. \textcolor{blue}{https://doi.org/10.1038/s42254-024-00703-6}

\bibitem{DE} J. A. Frieman, M. S. Turner, D. Huterer, ``Dark Energy and the Accelerating Universe", Annual Review of Astronomy and Astrophysics, 46, 385-432, 2008. \textcolor{blue}{https://doi.org/10.1146/annurev.astro.46.060407.145243}


\bibitem{dm1} R. W. Tucker, C. Wang, ``Dark matter gravitational interactions", Classical and Quantum Gravity, vol. 15, no. 4, p. 933, 1998. \textcolor{blue}{https://dx.doi.org/10.1088/0264-9381/15/4/015}

\bibitem{dm2} D. Clowe, M. Bradač, A. H. Gonzalez, \textit{et al.}, ``A Direct Empirical Proof of the Existence of Dark Matter*", The Astrophysical Journal, 648(2), L109, 2006. \textcolor{blue}{https://dx.doi.org/10.1086/508162}

\bibitem{dm3} C. Lage, G. R. Farrar, ``The bullet cluster is not a cosmological anomaly", Journal of Cosmology and Astroparticle Physics, 2015(02), 038–038, 2015. \textcolor{blue}{https://dx.doi.org/10.1088/1475-7516/2015/02/038}

\bibitem{bsm} P. Langacker, ``The Standard Model and Beyond (1st ed.)", CRC Press, 2010. \textcolor{blue}{https://doi.org/10.1201/b12328}

\bibitem{tor1} T. W. B. Kibble; Lorentz ``Invariance and the Gravitational Field", J. Math. Phys. 2, 212–221, 1961. \textcolor{blue} {https://doi.org/10.1063/1.1703702}

\bibitem{tor2}
F. W. Hehl, P. von der Heyde, G. D. Kerlick, \textit{et al.}, ``General relativity with spin and torsion: Foundations and prospects", Rev. Mod. Phys. 48(3), 393-416, 1976. \textcolor{blue}{https://doi.org/10.1103/RevModPhys.48.393}

\bibitem{tor3}
M. Shaposhnikov, A. Shkerin, I. Timiryasov, \textit{et al.}, ``Erratum: Einstein-Cartan Portal to Dark Matter [Phys. Rev. Lett. 126, 161301 (2021)]", Phys. Rev. Lett., 127(16), 169901, 2021. \textcolor{blue}{https://doi.org/10.1103/PhysRevLett.127.169901}

\bibitem{Einstein-Cartan} C. H. Nam,
``Probing dark gauge boson via Einstein-Cartan portal", Phys. Rev. D 105, 075015, 2022 \textcolor{blue}{https://doi.org/10.1103/PhysRevD.105.075015}

\bibitem{ilc1} T. Behnke, J. Brau, B. Foster, \textit{et al.}, ``The International Linear Collider Technical Design Report - Volume 1: Executive Summary", 2013. \textcolor{blue}{ 	
https://doi.org/10.48550/arXiv.1306.6327}



\bibitem{ilc2} H. Baer, T. Barklow, K. Fujii, \textit{et al.}, ``The International Linear Collider Technical Design Report - Volume 2: Physics", 2013. \textcolor{blue}{ 	
https://doi.org/10.48550/arXiv.1306.6352}


\bibitem{Extra-Gauge-bosons} M. Cvetic and S. Godfrey, “Discovery and identification of extra gauge bosons”. \textcolor{blue}{arXiv:hep-ph/9504216}

\bibitem{gaugeboson1} 
A. Leike, “The Phenomenology of extra neutral gauge bosons”, Phy. Rep. 317, 143, 1999. \textcolor{blue}{arXiv:hepph/9805494}

\bibitem{LR-symmetry11} 
M. Cvetic, P. Langacker, and B. Kayser, “Determination of g-R / g-L in left-right symmetric models at hadron colliders”, Phys. Rev. Lett. 68, 2871, 1992. \textcolor{blue}{https://doi.org/10.1103/PhysRevLett.68.2871}

\bibitem{Super-symmetry12} 
S. Dimopoulos and H. Georgi, “Softly Broken Supersymmetry And SU(5)”, Nucl. Phys. B 193, 150, 1981. \textcolor{blue}{https://doi.org/10.1016/0550-3213(81)90522-8}

\bibitem{bsmilc}
P. Bambade, T. Barklow, T. Behnke, \textit{et al.},  ``The International Linear Collider: A Global Project", 2019. \textcolor{blue}{https://arxiv.org/abs/1903.01629}

\bibitem{cms} CMS Collaboration,
``Search for resonant and nonresonant new phenomena in high-mass dilepton final states at $\sqrt{s}$ = 13 TeV", Journal of High Energy Physics (7), 2021. \textcolor{blue}{https://doi.org/10.1007/JHEP07(2021)208}

\bibitem{ATLAS}
ATLAS Collaboration, ``Search for high-mass dilepton resonances using 139 fb\(^{-1}\) of pp collision data collected at \(\sqrt{s} = 13\) TeV with the ATLAS detector", Physics Letters B, 796, 68–87, 2019. \textcolor{blue}{http://dx.doi.org/10.1016/j.physletb.2019.07.016}

\bibitem{zprimeCMSlowmass} CMS Collaboration,
Search for a Narrow Resonance Lighter than 200 GeV Decaying to a Pair of Muons in Proton-Proton Collisions at $\sqrt{s}$ = 13 TeV, Phys. Rev. Lett. 124, 131802 (2020).

\bibitem{whiz}
W. Kilian, T. Ohl, J. Reuter, ``WHIZARD—simulating multi-particle processes at LHC and ILC", The European Physical Journal C, 71(9), 2011. \textcolor{blue}{http://dx.doi.org/10.1140/epjc/s10052-011-1742-y}

\bibitem{pyth}
T. Sjöstrand, S. Mrenna, P. Skands, ``PYTHIA 6.4 physics and manual", Journal of High Energy Physics (05), 026–026, 2006. \textcolor{blue}{http://dx.doi.org/10.1088/1126-6708/2006/05/026}

\bibitem{delph}
J. de Favereau, C. Delaere, P. Demin, \textit{et al.}, ``DELPHES 3: a modular framework for fast simulation of a generic collider experiment", Journal of High Energy Physics, (2), 57, 2014. \textcolor{blue}{https://doi.org/10.1007/JHEP02(2014)057}

\bibitem{sig} G. Cowan, K. Cranmer, E. Gross, \textit{et al.},
``Asymptotic formulae for likelihood-based tests of new physics", Eur. Phys. J. C 71, 1554, 2011. \textcolor{blue}{https://doi.org/10.1140/epjc/s10052-011-1554-0}

\bibitem{CL1}
A. L. Read, ``Presentation of search results: the CLs technique", J. Phys. G: Nucl. Part. Phys., 28, 2693, 2002. \textcolor{blue}{https://doi.org/10.1088/0954-3899/28/10/313}

\bibitem{CL2}
T. Junk, ``Confidence level computation for combining searches with small statistics", ``Nuclear Instruments and Methods in Physics Research Section A: Accelerators, Spectrometers, Detectors and Associated Equipment", 434(2–3), 435-443, 1999. \textcolor{blue}{https://doi.org/10.1016/S0168-9002(99)00498-2}





\end{thebibliography}
\end{document}